# The ATA Digital Processing Requirements are Driven by RFI Concerns

G. R. Harp

## Abstract

As a new generation radio telescope, the Allen Telescope Array (ATA) is a prototype for the square kilometer array (SKA). Here we describe recently developed design constraints for the ATA digital signal processing chain as a case study for SKA processing. As radio frequency interference (RFI) becomes increasingly problematical for radio astronomy, radio telescopes must support a wide range of RFI mitigation strategies including online adaptive RFI nulling. We observe that the requirements for digital accuracy and control speed are not driven by astronomical imaging but by RFI. This can be understood from the fact that high dynamic range and digital precision is necessary to remove strong RFI signals from the weak astronomical background, and because RFI signals may change rapidly compared with celestial sources. We review and critique lines of reasoning that lead us to some of the design specifications for ATA digital processing, including these: beamformer coefficients must be specified with at least 1° precision and at least once per millisecond to enable flexible RFI excision.

## Introduction

Some of the worst sources of radio frequency interference (RFI) for radio astronomy are low earth orbit (LEO) satellites. Unlike ground-based RFI sources, there is no place on earth where we can hide from such satellites. Because of their low orbit, their angular position on the sky can vary faster than 1° per second. They broadcast strong signals that sometimes impinge on protected radio astronomical radio bands. For example, a recent study shows that Iridium satellites generate unwanted signals that are 25 dB above the detrimental level in the 1610.6-1613.8 MHz radio astronomy band.[*] Another study [*Ellingson, Hampson and Johnson*, 2003] finds RFI of unknown origin in the protected band centered at 1413 MHz. Such RFI presents a major challenge for astronomers attempting to reduce the damaging effects of these sources, and high speed calculations are required to simulate and remove their signals.

The square kilometer array (SKA) must face up to this challenge. Because RFI mitigation is so important for the SKA's success, it must be built into the design from the start [*Smolders and Hampson*, 2002, *Warr et al.*, 2002]. In this regard the Allen Telescope Array (ATA) provides a case study [*Welch and Dreher*, 2000]. Having just completed the detailed design, we report that ATA's digital signal processing requirements are driven by RFI mitigation. Indeed, we discovered that at current technology and funding levels we cannot build a system that is flexible enough to support all desired modes of RFI suppression.

---

[*] Leeheim Satellite Monitoring Station (2004), Report on the Monitoring Request on Iridium by the ECC chairman.

In this paper we use realistic simulations of active deterministic nulling of RFI from LEO satellites to quantify requirements for ATA signal processing. These simulations are then used to define requirements that enable adaptive nulling [*Barnbaum and Bradley*, 1998, *Ellingson and Hampson*, 2002, *Kesteven et al*., 2005], post-correlation image processing [*Leshem, van der Veen and Boonstra*, 2000], and reference antenna techniques [*Leshem, van der Veen and Boonstra*, 2000, *Mitchell and Robertson*, 2005] at the ATA. Indeed the practical application of *deterministic* nulling will be limited at the ATA due to the fact that the far out antenna sidelobe patterns are nearly impossible to calibrate [*Thompson*, 2003]. One case where deterministic nulling may be useful is when there exists a strong astronomical source such as a supernova remnant in the antenna primary beam but away from the synthetic beam maximum. Such a source makes a noise like contribution to the beamformer output and some experiments might benefit by nulling this out. Despite the fact that deterministic nulling will seldom be used, it still provides an accurate modeling tool because the fringe rotation rate due to satellite motion is greater than the rate of change in the antenna sidelobe gain pattern as the RFI source traverses the unknown sidelobe response.

## ATA Digital Processing and Simulations

The ATA is a privately-funded array currently under construction at Hat Creek Radio Observatory in northern California [*Welch and Dreher*, 2000]. It is being built in stages, first with 32 elements, then 206, then 350 elements for a total collecting area of about one hectare. The ATA data processing system has been developed and examined in several previous reports [*D'Addario*, 2001a; *Urry*, 2001; *DeBoer and Dreher*, 2003, *D'Addario*, 2001b; *D'Addario*, 2002, *Harp*, 2002]. After this processing, the signals are passed on to an imaging correlator [*Wright*, 2001; *Urry*, 2002; *Wright*, 2002; *Wright and Harp*, 2003] and to beamformers.

The signal processing leading to an ATA beamformer is conceptually depicted in Fig. 1 (an overview of radio interferometric signal processing can be found in [*Thompson, Moran and Swenson*, 2001]). Each antenna receives signals in two linear polarizations, X and Y. We display only the X path, the Y path is similar. The signal from each antenna is split into four copies which are downconverted using four independent local oscillators A, B, C and D. We follow signal path A. The signal passes through an analog anti-alias filter and is digitized at 155 MS/s (complex sampling). This signal is again split four ways, numbered 1-4. We follow path 1A1.

A delay $\tau_{1A1}$ is applied in the digital domain, with subsample accuracy afforded by an 8-tap FIR filter. The FIR filter additionally acts as a digital anti-alias filter and downsampler, outputting complex samples at the rate of 103 MS/s. The combination of analog and digital anti-alias filters is such that the useful bandwidth exceeds 90% of the Nyquist bandwidth in the 103 MS/s signal [*D'Addario*, 2002]. The FIR filter coefficients are chosen to put the delay center at the center of the antenna primary beam and performs the bulk of the beamforming. This signal multiplied by a single complex coefficient. The coefficient $c_{1A1}$ performs triple duty as it must a) correct the fringe rotation associated with $\tau_{1A1}$, b) steer the synthesized beam from the delay center to the desired position on the sky (usually within the antenna primary beam), and c) perform RFI mitigation like adaptive nulling, if any. In this context, it is convenient to think of $c_{1A1}$ as the product of three coefficients that accomplish these separate tasks. Finally, the signals from all antennas (or a preselected subset) are summed to give one of sixteen dual polarization beamformer outputs. Only 14 of these beamformers will be implemented as shown, the remaining 2 sets of signal paths are directed to imaging correlators.

In this paper we focus on requirements for digital control of $\tau$ and $c$ at the communication interface between two subsystems of the ATA. The first is the ATA control software (host), which is distributed over multiple computers and linked by a local area network. The second subsystem is called the IF processor, comprising everything to the right of the ADCs in Fig. 1. Because of the requirements for high speed processing, the IF processor is not implemented in general purpose hardware but with hundreds of custom-designed, field programmable gate array (FPGA) circuit boards. Because of the relative difficulty associated with reprogramming FPGAs, we keep the FPGA software simple and as flexible as possible. As we experiment with algorithms for calculating the beamformer coefficients with and without RFI mitigation, we choose to calculate these values on the host and then distribute them to the IF processor where they are applied. The information content of these data is quite high, and it is desirable to find a representation that compresses $\tau$ and $c$ to manageable rates. As we shall see, appropriate compression of these data is subtle.

To elucidate the information content of $\tau$ and $c$, we perform simulations with the currently proposed configuration of the ATA-350. For simplicity, the observation source is placed at telescope zenith. The RFI source is assumed to be in polar orbit on a path that passes directly over the observatory, and the simulations put the satellite in the vicinity of elevation 50º. These calculations use techniques we have developed specifically for the ATA, that generate wide frequency band nulls in a single-tap beamformer [*Harp*, 2002; *Harp and Ellingson*, 2003]. In line with our comments above, we ignore antenna primary beam variations since the main lobe width is about 100 times larger than that of the synthetic beam pattern. Hence primary beam variations will not impact the calculated rates of change of the beamformer coefficients.

## Specifying Delay and Complex Gain

We expect that RFI mitigation will be very important for the success of the ATA and design the data processing chain to be as flexible as possible for real-time RFI removal. This is especially important for devices that rely on beamformer outputs, since post-correlation techniques for RFI removal are not applicable in this case. Although the delays are not used directly to perform RFI mitigation, we include them in our analysis since a) they are required for control of the beamformers and b) they provide a mathematical starting point for calculation of the complex gain.

With this in mind, we notice that some RFI sources are ground-stationary while others (satellites, airplanes) move across the sky at varying speeds. We begin by examining the requirements for tracking and forming a beam on a moving RFI source, on the assumption that RFI removal can only be more difficult than this. The requirement is set, somewhat arbitrarily, that we can track RFI sources moving at least as fast as a 350 km altitude low earth orbit (LEO) satellite. This includes almost all present or planned LEO's but is not fast enough to follow the international space station or an airplane flying directly over the site.

The linear velocity of a LEO is nearly independent of altitude and can be approximated by equating the centripetal force with the force of gravity, giving $v \approx \sqrt{rg} \approx 8$ km/s. If we approximate the earth's surface with a plane, we can estimate the angular velocity of a satellite (as viewed from the ground) in an overhead pass as a function of elevation angle $\theta$, and altitude $h$:

$$\dot{\theta} = \frac{v}{h}\sin^2\theta, \quad \dot{\theta}_{\max} = 1.3^{\circ}/\mathrm{s}. \qquad [1]$$

The interference signal path length difference between the array center and an antenna at radius $d$ is $p = d\cos\theta$. The ATA antennas are arrayed over a 1 km x 1 km area. For an antenna at the extreme edge of the array ($d$ = 500 m from center) this yields,

$$\tau_{\max} = \frac{d}{c}\cos\theta = 1.6 \times 10^{-6}\cos\theta \text{ seconds.} \qquad [2]$$

From this we can estimate the maximal delay and maximal rate of change of delay

$$\begin{aligned} \tau_{\max} &\sim 4\frac{d}{c} = 7 \times 10^{-6}\ \mathrm{s} = 1000 \text{ samples, and} \\ \dot{\tau}_{\max} &\sim \frac{dv}{ch} = 4 \times 10^{-8} = 6 \text{ samples / s.} \end{aligned} \qquad [3]$$

where the factor of 4 in the equation for $\tau_{\max}$ comes from A) a conservative estimate of the variation in fiber optic cable lengths from antenna to antenna (factor of 2) and B) the fact that the set delay must be positive (another factor of 2). The conversion to samples uses the ATA IF sampling rate of 155 MHz. We require the digital delay be maintained to an accuracy of ±0.1 sample [*D'Addario*, 2001a]. This means that if we express $\tau$ directly at each moment in time, we require an update rate of 60 Hz and at least 14-bit precision or 840 bits per second to the IF processor.

After the delay, each signal path $j$ is multiplied by a single complex coefficient, $c_j$ (Fig. 1), also represented as an amplitude $\alpha_j$ and phase $\varphi_j$:

$$c_j = \alpha_j \exp(-i\varphi_j). \qquad [4]$$

The phase parameter $\varphi$ steers the phased array beam to any position on the sky independent of the delay setting. This ability is required because multiple beams will share the same delay (see Fig. 1). Hence we pursue a derivation parallel to the previous section, assuming that we must align our signals using phase only.

For a given frequency *f*, the relationship between delay and phase is linear

$$\varphi(t) = 2\pi f \tau(t). \qquad [5]$$

Using $f_{\max}$ = 11.2 GHz[†], we can immediately conclude that

$$\begin{aligned}
\varphi_{\max} &\sim 2\pi f_{\max}\frac{d}{c} = 1\text{ x }10^5\,\text{rad} = 7\text{ x }10^6\,\text{deg},\\
\dot{\varphi}_{\max} &\sim 2\pi f_{\max}\frac{dv}{ch} = 3\text{ x }10^3\,\text{rad / s} = 2\text{ x }10^5\,\text{deg / s},\\
\ddot{\varphi}_{\max} &\sim 4\pi f_{\max}\frac{dv^2}{ch^2} = 100\text{ rad / s}^2 = 7\text{ x }10^3\,\text{deg / s}^2\text{, and}\\
\dddot{\varphi}_{\max} &\sim 12\pi f_{\max}\frac{dv^3}{ch^3} = 10\text{ rad / s}^3 = 5\text{ x }10^2\,\text{deg / s}^3.
\end{aligned} \qquad [6]$$

Note that $\varphi$ need only be specified modulo $2\pi$.

For SETI observations where the target is unresolved and signal to noise is the dominant concern, $\alpha$ is unity. But most other astronomical observations demand the ability to taper the phased array beam. In the area of RFI mitigation, it is possible to perform RFI cancellation without controlling $\alpha$ [Kogan, 2005, Fridman, 2005], but it is generally more convenient to have $\alpha$ control. In particular, setting $\alpha$ = 0 is how we plan to implement RFI blanking which is particularly useful for mitigating pulsed radar signals [*Niamsuwan, Johnson and Ellingson*, 2005].

We have set the requirements that $\varphi$ and $\alpha$ must maintain digital accuracies of ±1° and ±1%, respectively. This choice was made based on expected phase and amplitude calibration errors as described in [*Wright*, 2001]. Similar to [*Torres*, 2000], the beamformers will be calibrated using real-time output from the imaging correlator. Using self-calibration there are sufficient number of relatively strong sources to achieve at least 5º phase error and 10% amplitude error without repointing the antennas.

Our choice of digital accuracy is further justified by the simulations of Fig. 2. Here we display the frequency dependence of the ATA-350 synthetic beam pattern along the direction toward a point-like or spatially unresolved RFI source. For this beam pattern, the source is at zenith and RFI is at 50º elevation. The observation frequency is 1420 MHz, and a 3 MHz wide bandwidth null centered at this frequency was generated by placing 9 equispaced single-frequency nulls over the desired range. The algorithm described in [*Harp*, 2002] was iterated until each point null had reached a depth at least 50 dB below the synthetic beam maximum. The output of these calculations is a set of coefficients $c_j$, whereas the delay values are determined by the position of the source. Following this calculation, Gaussian-distributed random noise was added to $\varphi$ or $\alpha$ for each antenna. These simulations help us quantify the amount of inaccuracy we can tolerate in the setting of $c$, whether that inaccuracy comes from imprecise knowledge of the analog components or from finite digital precision.

---

[†] The requirements for phase accuracy are in some ways more stringent than for delay accuracy. The delay accuracy is calculated using a frequency related to the total bandwidth of the observation (in this case, 50 MHz at the band edge). The phase, on the other hand, is applied at the sky frequency (11.2 GHz), so a much greater dynamic range is required.

For ideal coefficients we notice that the null regions are everywhere below -50 dB relative to the synthetic beam maximum. For 5º phase or 10% amplitude noise associated with calibration errors there is already a significant degradation in the effectiveness of the pattern null. To prevent further degradation coefficient values must be controlled with an accuracy higher than this. The simulations with 1° phase or 1% amplitude error suggest the negative impact of the digital accuracy alone. These results are in good agreement with those of Thompson [*Thompson*, 2003], who performed simulations of 1-D arrays. Similar issues are explored in [*Smolders and Hampson*, 2002] and in [*Kogan*, 2005].

With a digital accuracy of 1°, a direct expression of $\varphi$ as a function of time (Eq. [6]) requires 2 x $10^5$ updates per second with 8-bit accuracy, for a total of 1.6 Mbit per second transmitted to the IF processor. This is a very high data rate. With 350 antennas and 16 independent beams, the aggregate data rate is 9 Gbit / s. It is desirable to consider more compact representations that can reduce the communication bandwidth. We do this in the next section.

## IF Processor Update Rate

In the last section we concluded that the complex coefficients (Eq. [4]) dominate delay (Eq. [2]) in terms of information content. Therefore we concentrate our attention on these coefficients and how to represent them efficiently. The form of (Eq. [4]) suggests a variety of data representations including a discrete Fourier transform (DFT). Our initial approach represents the amplitude and phase with Taylor's series, e.g.

$$\varphi(t) = \varphi_0 + \varphi_1 t + \varphi_2 t^2 + ... \qquad [7]$$

If all we want to do is track a 350 km altitude LEO, Eq. [6] gives a straightforward estimation of the minimum time a single set of expansion coefficients shall remain accurate depending on the number of terms in the Taylor's series. For $\Delta\varphi = 1°$:

$$\begin{aligned} &1\,\text{Term}: \quad t_{\max} \sim \frac{\Delta\varphi}{\dot{\varphi}_{\max}} = 7 \text{ x } 10^{-6} \text{ s} \\ &2\,\text{Terms}: \quad t_{\max} \sim \sqrt{\frac{2\Delta\varphi}{\ddot{\varphi}_{\max}}} = 2 \text{ x } 10^{-2} \text{ s} \qquad [8] \\ &3\,\text{Terms}: \quad t_{\max} \sim \sqrt[3]{\frac{6\Delta\varphi}{\dddot{\varphi}_{\max}}} = 0.2 \text{ s.} \end{aligned}$$

This appears quite promising. With a 3-term Taylor's series (Eq. [7]) we can specify $\varphi$ for up to 0.2 seconds, as long as we maintain sufficient precision in the coefficients. This corresponds to a data rate of less than 1 kbit / second, compared to 3.2 Mbit / s above. Similar analysis can be done for the delay and amplitude.

This analysis suffices for tracking a LEO, but it is incorrect for nulling the RFI coming from a LEO. We discovered this by performing simulations of time-dependant beamformer coefficients. In Fig. 3 we show calculations of the amplitude and phase of one randomly selected antenna as the array performs observations at 1420 MHz and nulling a 350 km LEO. The calculations are similar to those for Fig. 2 and place a deterministic 10 MHz bandwidth null at the position of the LEO.

In Fig. 3 we are immediately struck by the fact that the amplitude and phase are not smooth and appear to have discontinuities. A close analysis shows that these are not true discontinuities, and that Eq. [6] is close to predicting the maximal observed value of $\dot{\varphi}_{\max}$, but it greatly underestimates $\ddot{\varphi}_{\max}$ and $\dddot{\varphi}_{\max}$. Guided by this result, we restrict ourselves to 2-term Taylor's series expansions in the next section. For the moment we abandon the analytic approach and derive the properties of the $c_j$ from realistic simulations.

**Simulations of Nulling at the ATA-350**

The worst-case scenario arises when the LEO has the lowest altitude and the observation frequency is highest; in our case the extremal parameters are 350 km, and 11.2 GHz, respectively. Simulations were performed using these parameters, observing at zenith and the LEO at elevation angle of 50º, and with a 10 MHz bandwidth null. These look qualitatively similar to Fig. 3, except that the variations occur on a shorter time scale. To better characterize these variations we take their Fourier transform and present the frequency power spectra of $\alpha$, $\varphi$, Re($c$), and Im($c$) in Fig. 4. This points to the first of our conclusions: Contrary to our intuition, $\alpha, \varphi$ is not an advantageous representation for $c$. Because evaluation of $c$ from $\alpha$ and $\varphi$ requires trigonometric functions whereas Re($c$), Im($c$) does not, we choose the latter representation. This result is intrinsically linked to the ATA's needs for RFI mitigation; a recently designed correlator for VLBI observations [*Carlson et al*., 1999] which are less sensitive to RFI chooses the phase and amplitude representation.

In Fig. 4 there is a clear frequency cutoff, $\nu_{\max}$, in Re($c$) and Im($c$) near 600 Hz. Where does this cutoff come from? To answer this question we performed simulations with various observation frequencies, $f$, and array sizes, $d$. The array size was varied by taking the antenna positions and scaling their separation by a uniform constant. We find the cutoff frequency is proportional to $d \cdot f$. Comparing with the second line of Eq. [6], we now recognize the relationship between $\dot{\varphi}_{\max}$ and $\nu_{\max}$. The highest frequency in the Fourier expansion of $c$ comes from the longest baseline in the array. For time passage $\nu_{\max}^{-1}$, the maximal phase change between the two extremal antennas is $2\pi$. Since the longest baseline is 2*d*, a derivation similar to Eq. [6] gives

$$\nu = \frac{2d\,f\,v}{ch}\sin^3(\theta), \quad \text{and} \quad \nu_{\max} = 900 \text{ Hz.} \qquad [9]$$

The reason we obtained $\nu_{\max} \approx$ 600 Hz in our calculation is that the calculation was performed at lower elevation than zenith. We note that Equation [9] is in agreement with a similar analysis performed by Thompson [*Thompson*, 2003].

One might ask whether the bandwidth of the null has any impact on the power spectrum of *c*? We investigated this question by performing simulations with different bandwidths including some with the minimum possible bandwidth. We find that frequency bandwidth has no significant effect on $\nu_{\max}$, although amplitude variations are reduced for very narrow bandwidths.

**Discussion**

Equation [9] could have been derived directly following an analytical approach without resorting to RFI simulations. However, without simulations it is easy to be lead astray with regard to the time dependence of the beamformer coefficients, *c*. We emphasize that when tracking a *single* source, Eq. [6] correctly characterizes the time dependence of the $c_j$. However, when tracking one source and simultaneously nulling another, each $c_j$ is not well represented by a single sinusoidally varying term and Eq. [6] breaks down for derivatives greater than the first.

This insight also helps us to understand why the null bandwidth had little impact on the power spectrum of *c*. No matter what bandwidth is chosen, the time dependence of the $c_j$ is still dominated by period of the longest baseline.

For the ATA design, we conclude that the beamformer coefficients, *c*, are quite unpredictable. We choose to express them with two-term Taylor's series expansions:

$$\begin{aligned} c &= a + ib, \text{ with} \\ a &= a_0 + a_1 t \\ b &= b_0 + b_1 t \end{aligned} \qquad [10]$$

where $a_0, a_1, b_0,$ and $b_1$ are real constants updated at 1 kHz. This coefficient rate is comparable to that required for space-based VLBI processing which doesn't demand real-time RFI mitigation [*Carlson et al.*, 1999]. The cost to radio astronomy of a polluted RF environment is enormous.

Until now our discussion has focused on real-time RFI mitigation in the beamformers. In the ATA imager (correlator) the same real-time techniques can be applied, although they become M-times harder where M is the number of pixels in the correlator image. Because the correlator preserves the complex nature of the astronomical data, it is convenient to perform RFI mitigation after the fact. This is possible provided the integration time (dump time) of the correlator is sufficiently short to prevent loss of phase information from the RFI signal. From Eq. [9] we immediately find that for a 350 km LEO, the integration period must be ~1 ms. Compare this with the 1 minute correlator dump time at the recently decommissioned Berkeley Illinois Maryland Array [*Welch et al.*, 1977] which operated at 10x higher frequencies but did not have to deal with RFI.

The 1 ms dump time mentioned above will not be achieved in the ATA imager. Because shorter integration periods result in proportionally larger data outflow from the correlator, it was decided that the ATA imager could not support periods shorter than 10 ms. This may limit our capabilities for off line RFI mitigation in astronomical images.

As a point of reference, it is interesting to compare the ATA requirements with those for the Expanded Very Large Array (EVLA), an upgrade to a well-established radio interferometer in Socorro New Mexico. Currently under construction, the EVLA must tackle the same RFI issues as does the ATA. The design requirements for the EVLA correlator demand a dump time of less than 100 ms [*Benson and Owen*, 2000], which is comparable to the ATA imager. Meanwhile, simulations of post-correlation RFI mitigation at the EVLA [*Perley and Cornwell*, 2003] confirm that to be successful, correlations must be “sampled quickly enough to prevent phase smearing.” In the configuration most similar to the ATA (D-array), Perley and Cornwell find a required dump rate of 21 ms at 10 GHz under typical conditions. Again, this is comparable to the values found here.

Using Eq. [9] we project this understanding to the requirements of the SKA. Using a maximum baseline length of 300 km and a maximum frequency of 20 GHz, we find that the SKA beamformer coefficients have $\nu_{max}$ = 0.6 MHz. This overestimates the actual required rate since the SKA will probably be constructed with multiple “stations” of clustered antennas. RFI mitigation can be performed station-by-station before final combination of signals [*Thompson*, 2003], hence the update rate may be similar to that of the ATA. This is especially important for the SKA imager, which must dump its data product on the same time scale. Given computational improvements between now and when SKA is built, it may be possible to dump the SKA correlator on a 1 ms timescale, but not on a 1 μs time scale.

## Conclusion

It is a sad reflection of the problems facing radio astronomy to note that the requirements for digital processing at the ATA have nothing to do with tracking astronomical sources and everything to do with RFI. Although some RFI (consumer electronics, cell phones, broadcast radio and TV) can be mitigated by choosing a remote location, it is impossible to conceal your interferometer from LEO satellites whose selling point is that they provide coverage of the entire earth’s surface. Our best hope for seeing through these interferers is with adaptive RFI excision either in real time or after the fact. This excision is possible only if the data processing system can keep up with the high speeds necessary to track and remove these signals.

While designing the signal processing system we examined various approaches for estimating the update speed necessary to mitigate interference from LEO’s. We find that the beamformer coefficient update rate is surprisingly high (1 kHz) and there is no obvious way to further compress these coefficients for transmission to the data processing hardware. At the SKA, our analysis suggests that similar update rates will be required.

## Acknowledgements

I am grateful to all of my amazing colleagues on the ATA project as the ideas described here were worked out among a closely knit group, including G. C. Bower, L. R. D'Addario, M. M. Davis, D. R. DeBoer, J. W. Dreher, G. Girmay-Kaleta, W. L. Urry, and M. C. H. Wright. This research was supported by the SETI Institute.

**Figures**

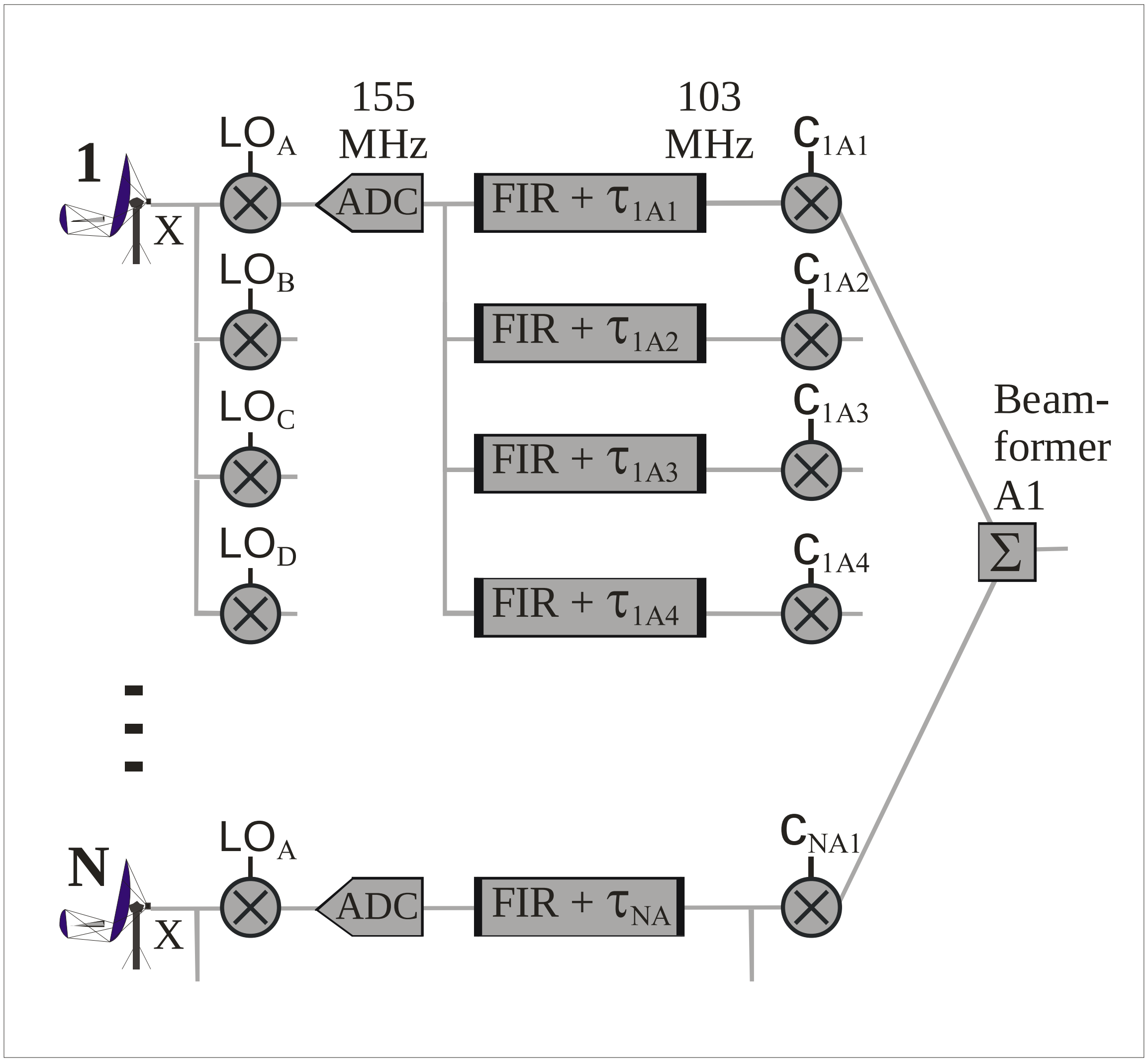

Figure 1: A schematic diagram of the signal processing leading to the ATA single-tap beamformer. The RF signal from each antenna is downconverted, delayed and multiplied by a single complex coefficient before summing to give a single-pixel beam.

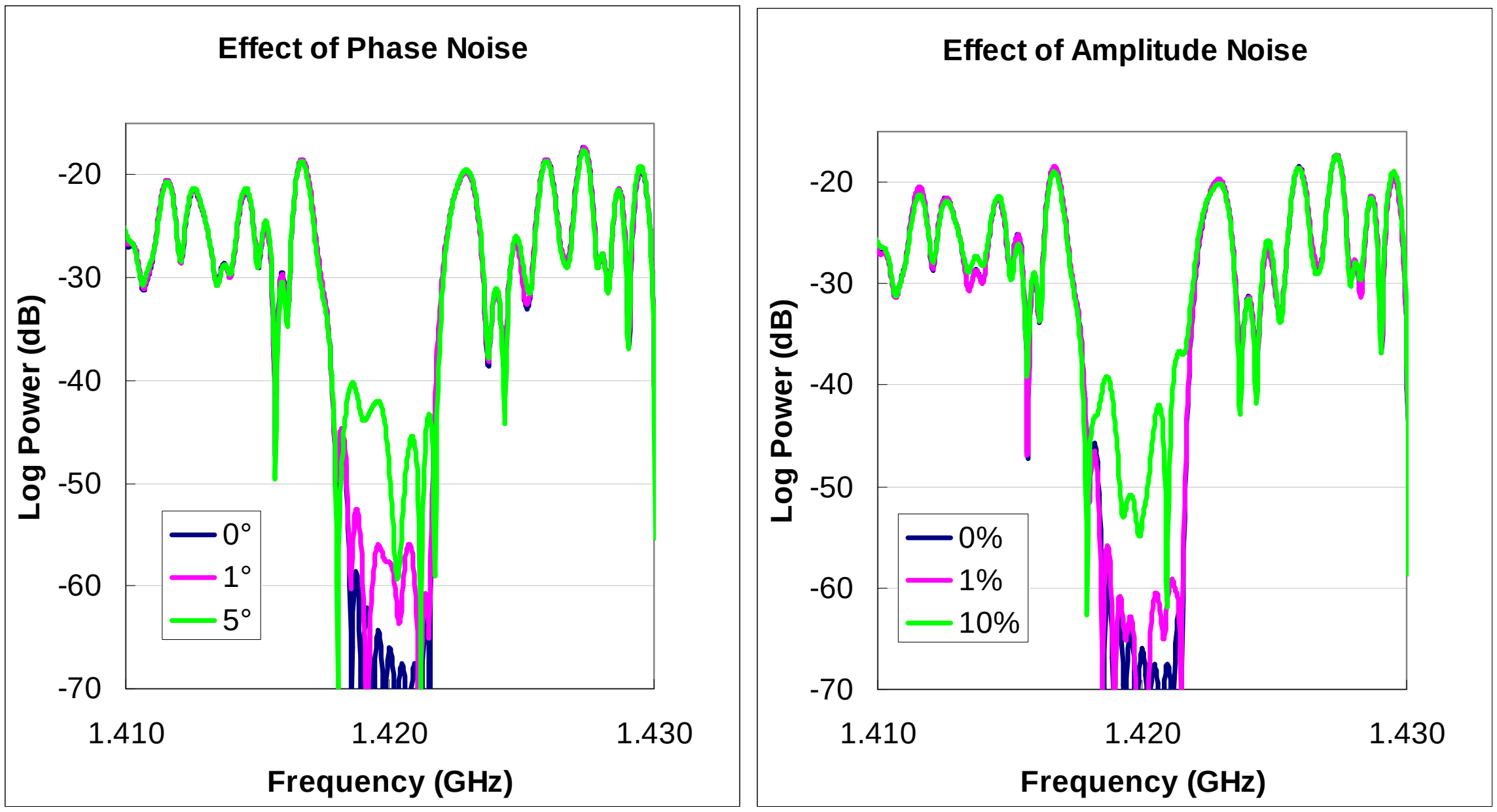

Figure 2: Simulations of a wide bandwidth null directed toward an unresolved RFI source. Here we plot the synthetic beam pattern in the direction of the RFI as a function of frequency. The unperturbed null (0° or 0%) is 3 MHz wide and centered at 1.42 GHz. The beamformer coefficients are then perturbed with Gaussian-distributed phase (left) or amplitude (right) noise, with RMS variation as indicated in the legend. As expected, increasing noise leads to a reduction in effectiveness of the null.

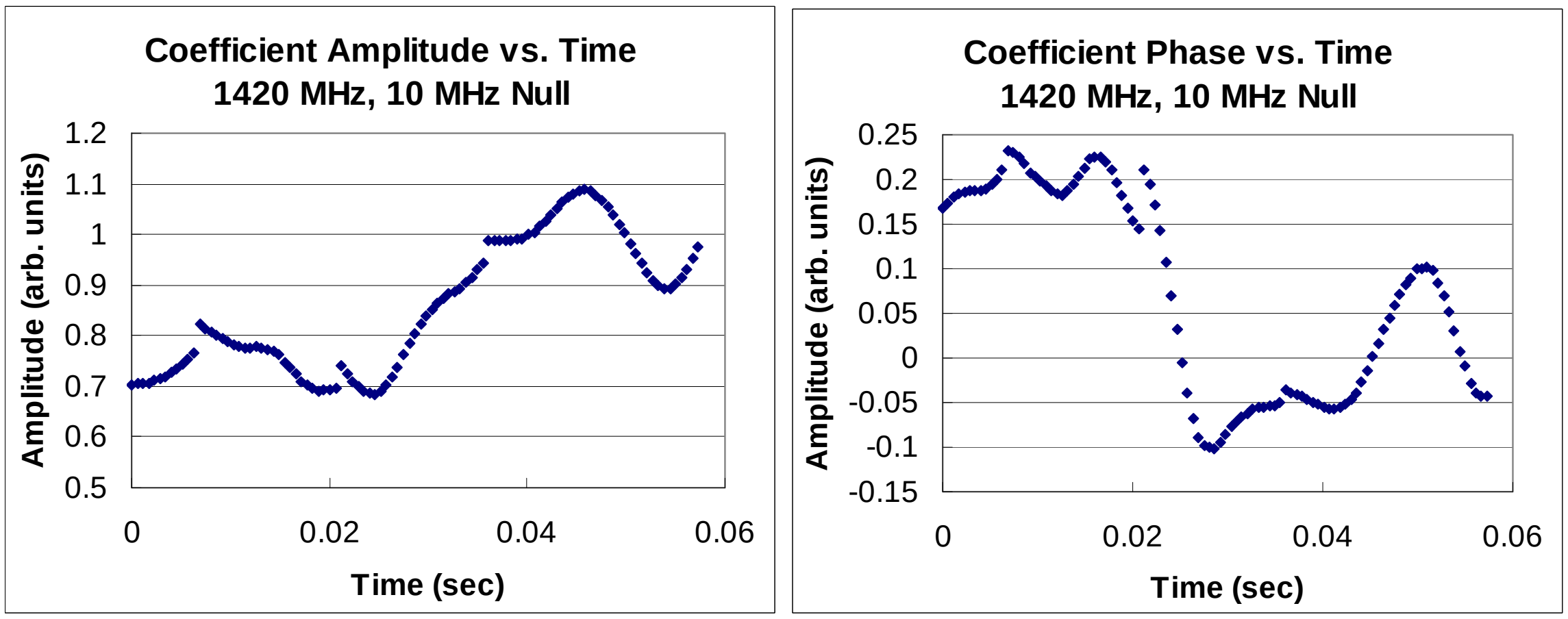

Figure 3: Simulations of the amplitude and phase of a randomly chosen beamformer coefficient, $c_j$, while tracking an astronomical source and placing a 10 MHz bandwidth null on a moving satellite. Surprisingly, we observe frequent near-discontinuities in the coefficient value. Such discontinuities make compression of the coefficient data difficult.

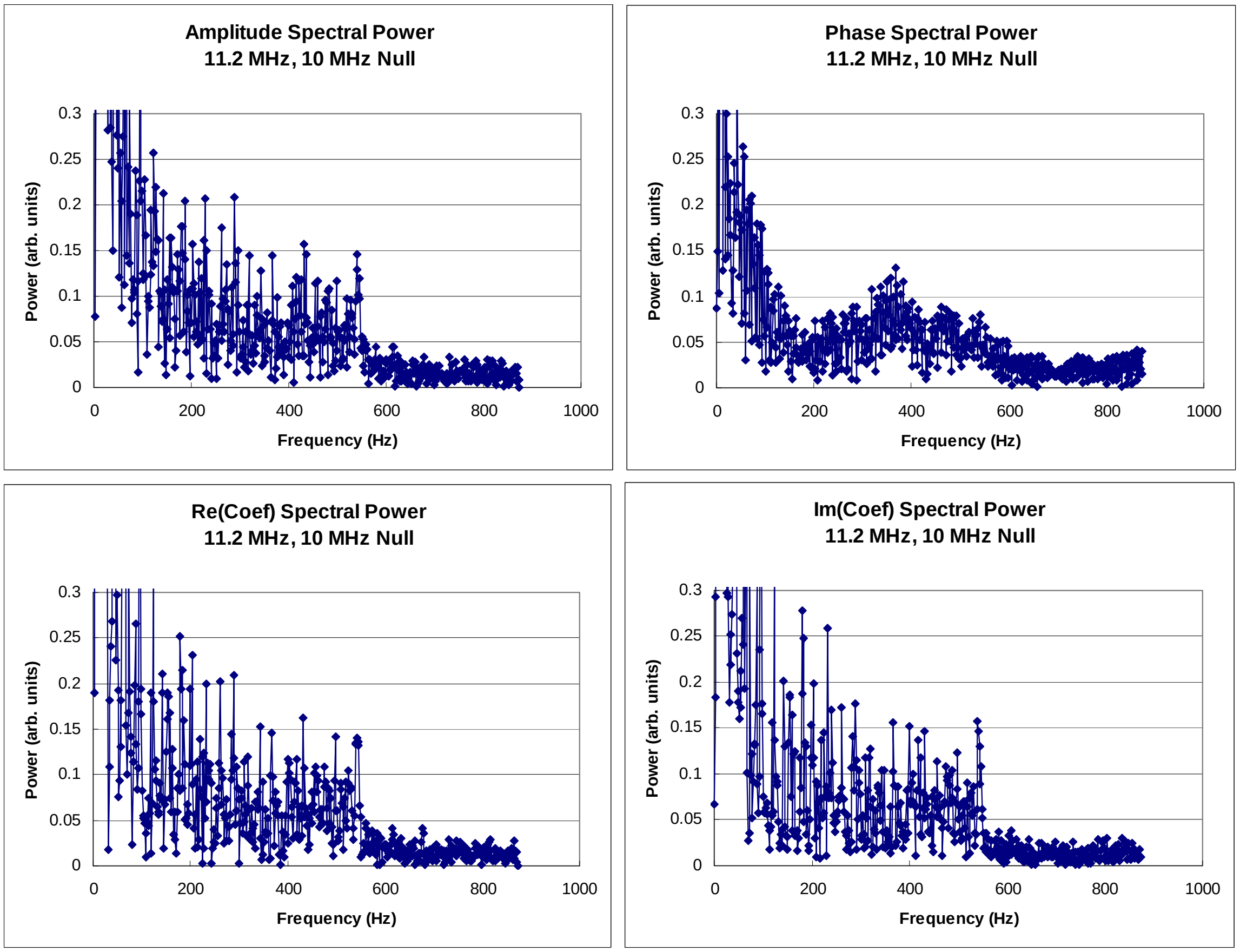

Figure 4: Power spectra of the beamformer coefficient, $c_j$, for a randomly chosen antenna. These simulations were performed using an observation frequency of 11.2 GHz and a 10 MHz RFI null. The sharp frequency cutoff in these spectra arises from the physical cutoff at the edge of the antenna array.